\documentclass{JAC2003}

\usepackage{graphicx}
\usepackage{booktabs}

\begin{document}
\title{E-cloud map formalism: an analytical expression for quadratic coefficient}
\author{T. Demma, INFN-LNF, Frascati (Italy),\\
S. Petracca, A. Stabile, University of Sannio, Benevento (Italy) \& INFN Salerno, Italy.}
\maketitle

\begin{abstract}
The evolution of the electron density during  electron cloud formation can be reproduced using a bunch-to-bunch iterative map formalism. The reliability of this formalism has been proved for RHIC \cite{Iriso} and LHC \cite{noi}. The linear coefficient has a good theoretical framework, while quadratic coefficient has been proved only by fitting the results of compute-intensive electron cloud simulations. In this communication we derive an analytic expression for the quadratic map coefficient. The comparison of the theoretical estimate with the  simulations  results shows a good agreement for a wide range of bunch population.
\end{abstract}


\maketitle

\section{INTRODUCTION}
In \cite{Iriso} it has been shown that, the evolution of the
electron cloud density can bedescribed 
introducing a quadratic map of the form:
\begin{eqnarray}\label{map}
n_{m+1}\,=\,\alpha\,n_m+\beta\,{n_m}^2
\end{eqnarray}
where $n_m+1$  and   $n_m$ are the average densities of electrons between two successive
bunches. The coefficients $\alpha$ and $\beta$ are
extrapolated from simulations and are functions of the beam
parameters and of the beam pipe characteristics. An analytic
expression for the linear map coefficient that describes electron
cloud behavior from first principles has been derived for straight sections of RHIC \cite{Iriso2}. In this paper we find
an analytical expression  the quadratic term coefficient.
We consider $N_{el,m}$ quasi-stationary electrons gaussian-like distributed in
the transverse cross-section of the beam pipe. The bunch $m+1$ accelerates the $N_{el,m}$
electrons initially at rest to an energy $\mathcal{E}_g$. After the first electrons-
wall collision two new jets are created: the backscattered electrons
with energy $\mathcal{E}_g$  and the "true
secondaries" (with energy $\mathcal{E}_0\,\sim\,5\,eV$).

The sum of these jets gives the number of
surviving electrons $N_{el,m+1}$, then one gets the linear coefficient
\begin{eqnarray}\label{linear_coeff}
\alpha\,=\,\frac{N_{el,m+1}}{N_{el,m}}
\end{eqnarray}
In the next section we compute the  quadratic term coefficient $\beta$
when the saturation condition of the electron cloud is obtained . Once calculated saturation we pass to estimate theoretically the coefficient $\beta$. We compare our results  with the  outcomes of numerical simulations
obtained using ECLOUD \cite{ecloud}. In the Table \ref{tab1} we report all parameters used for our calculations.
\begin{table}[hbt]
	\begin{center}
	\caption{Input parameters for analytical estimate and ECLOUD simulations.}
	\begin{tabular}{lcc}
		\hline
		\hline
		Parameter & Unit & Value \\
		\hline
    Beam pipe radius $b$ & m & $.045$ \\
    Beam size $a$ & m & $.002$ \\	
    Bunch spacing $s_b$ & m & $1.2$\\
		Bunch length $h$ & m & $.013$\\
    Energy for $\delta_{max}$ $\mathcal{E}_{0,max}$ & eV & $300$\\
    Energy width for secondary $e^-$ & eV & -\\
		Number of particles per bunch $N_b$ & $10^{10}$& $4\,\div\,9$\\
		Secondary emission yield (max)  $\delta_{max}$ & - & $1.7$\\
    Secondary emission yield ($\mathcal{E}\rightarrow0$)  & - & $.5$\\
    \hline
		\hline
	\end{tabular}
\label{tab1}
\end{center}
\end{table}

\section{Steady-state: electronic density of saturation}
\begin{figure}[htbp]
  \centering
  \includegraphics*[width=65mm]{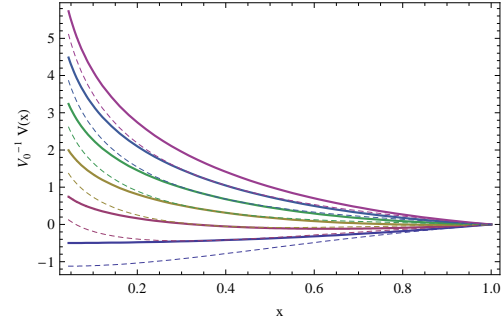}
  \caption{Plot of $V_0^{-1}V(x)$, (\ref{radialpotential}) and (\ref{potentialuniform})), in the case of uniform (solid lines) and gaussian (dashed lines) electronic distribution for $g\,=\,0\,\div\,2$, $\tilde{a}\,=\,.04$, $\tilde{r}_0\,=\,0$, $\tilde{\sigma}\,=\,.3$.}
  \label{plotpotential}
\end{figure}

\begin{figure}[htb]
  \centering
  \includegraphics*[width=65mm]{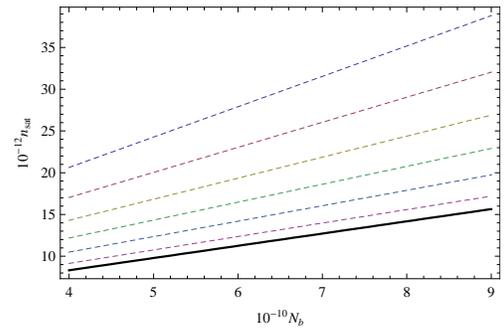}
  \caption{Plot of electronic densities of saturation $n_{sat}$ vs $N_b$, (\ref{nsatgauss}) and (\ref{nsatunif})), withf uniform (solid line) and gaussian (dashed lines) electronic distribution for $\tilde{a}\,=\,0.04$, $\tilde{r}_0\,=\,0$, $\tilde{\sigma}\,=\,0.3$ and $p\,=\,2\,\div\,3$.}
  \label{plotdensitysaturation}
\end{figure}

In the chamber we have two groups of electrons belonging
to cloud: primary photo-electrons generated by the synchrotron
radiation photons and secondary electrons generated by the beam
induced multi-pactoring. Electrons in the first group generated at
the beam pipe wall   interact with the parent
bunch and  are accelerated   to the velocity given by:
$v/c\,=\,2\bar{N}_br_e/b$, where $r_e$ is the classical electron
radius and $\bar{N}_b$ is the effective value of bunch population and
\begin{eqnarray}
\bar{N}_b\,=\,\frac{h}{h+s_b}N_b
\end{eqnarray}
$s_b$ ibeing the bunch spacing and $h$  the length of bunch.
Electrons in the second group, generally, miss the parent bunch and move from the
beam pipe wall with the velocity given by: $v/c\,=\,\sqrt{2\mathcal{E}_0/mc^2}$, $\mathcal{E}_0$ being the average energy of the secondary electrons, until
the next bunch arrives. 
The process of thecloud formation depends, respectively, on two parameters:
\begin{equation}
k\,=\,\frac{2\bar{N}_br_eh}{b^2}
\end{equation}
\begin{equation}
\xi\,=\,\frac{h}{b}\sqrt{\frac{2\mathcal{E}_0}{mc^2}}
\end{equation}
The second one is the distance (in units of $b$) passed by electrons of each group before the next bunch arrives. At low currents, $k\,<<\,1$, each electron interacts with many bunches before it reaches the opposite wall. In the opposite extreme case, $k\,>\,2$, all electrons go wall to wall in one bunch spacing. The transition to the second regime occurs when $k\,\sim\,1$ .
The density of the secondary electrons grows until the space-charge potential energy of the secondary electrons is lower than $\mathcal{E}_0$.
The saturation condition can be obtained by requiring that the potential barrier is greater than electron energy in the point $r/b\,=\,1-\xi$
\begin{equation}\label{condition}
-e\,V(1-\xi)\,\sim\,\mathcal{E}_0
\end{equation}
where $V$ is the electric potential generated by the bunch and the electron cloud. To calculate the electric potential we assume that  
our system is composed by a chamber with radius $b$, a bunch with radius $a$ and length $h$, an electron cloud with density $\rho$. We consider the following electron distribution :
\begin{equation}\label{radialdensity_1}
\rho(r)\,=\,\rho_0e^{\displaystyle{-\frac{(r-r_0)^2}{2\sigma^2}}}
\end{equation}
where $\rho_0$ is fixed by the condition
\begin{equation}
2\pi h\int_a^b\rho(r)rdr\,=\,-N_{el}\,e
\end{equation}
and $N_{el}$ is the total number of electrons in the volume $\pi h(b^2-a^2)$. The electric potential $V(r)$, defined by the condition $V(b)=0$ is:
\begin{equation}\label{radialpotential}
V(r) = -V_0 \left[g\ln x+\frac{G(x)}{F(1)}\right],
\end{equation}
where $F(x)\,=\,\int_{\tilde{a}}^x\,exp(-
(\tilde{y}-\tilde{r}_0)^2/2\tilde{\sigma}^2)y\,dy$, $G(x)\,=\,\int_x^1\,F(y)/ydy$, $g\,=\,\bar{N}_b/N_{el}$, $V_0\,=\,N_{el}\,e/2\pi\epsilon_0h$  and $x\,=\,r/b$, $\tilde{a}\,=\,a/b$, $\tilde{r}_0\,=\,r_0/b$, $\tilde{\sigma}\,=\,\sigma/b$. We note that if $\sigma\,>>\,b$ (or $\tilde{\sigma}\,>>\,1$) and $r_0\,=\,0$ we obtain the uniform electron cloud and with $a\,\rightarrow\,0$ we must neglect the radial dimension of bunch with respect to that one of electron cloud. In  this case equation (\ref{radialpotential}) gives
\begin{equation}\label{potentialuniform}
V(r)\,=\,-V_0\biggl[g\ln x+\frac{1-x^2}{2}\biggr]
\end{equation}
Obviously the potentials depend on $g$, the ratio of the densities
of the beam and of the cloud averaged over the beam pipe
cross-section. In FIG. \ref{plotpotential} we report the spatial
behavior of two potentials. The potential (\ref{potentialuniform})
has minimum at $r\,=\,r_m\,=\,b\sqrt{g}$ and is monotonic for
$g\,>\,1$ within the beam pipe. For $g\,<\,1$ it has minimum at
the distance $r_m\,<\,b$, and the condition $g\,=\,1$ defines the
maximum density. this is the well known condition of the
neutrality. The condition formulated in this form is, actually,
independent of the form of distribution. Similar behavior is found
also for the gaussian distribution density and is compared with
respect to previous one (FIG. \ref{plotpotential}).
By imposing the condition (\ref{condition}) we find the critical number (saturation condition) of
electrons in the chamber
\begin{equation}
N_{el,sat}\,=\,\frac{2\pi\epsilon_0hF(1)\mathcal{E}_0}{e^2G(1-\xi)}-\frac{
F(1)\ln(1-\xi)}{G(1-\xi)}\bar{N}_b
\end{equation}
while the average density of saturation is found by assuming that electrons are confined in a cylindrical shell with inner radius  $a$ and external radius  $r_0+p\,\sigma$ where $p$ is a free parameter. So
\begin{equation}\label{nsatgauss}
n_{sat}\,=\,\frac{N_{el,sat}}{\pi h b^2 [(\tilde{r}_0+p\,\tilde{\sigma})^2-\tilde{a}^2]}
\end{equation}
where $p$ is a free parameter. For a uniform electron cloud distribution  we find the  saturation density
\begin{equation}\label{nsatunif}
\bar{n}_{sat}\,=\,\frac{\bar{N}_{el,sat}}{\pi h b^2 [1-\tilde{a}^2]}
\end{equation}
In the FIG. \ref{plotdensitysaturation} we show the behavior of saturation density  (\ref{nsatgauss}) and (\ref{nsatunif}). It is obvious for a gaussian distribution   we get a estimate of density saturation greater than that of a uniform distribution. In fact, the same number of electrons occupies a smaller volume (due to the Gaussian distribution).

\section{Analytical determination of coefficients}
\begin{figure}[htb]
  \centering
  \includegraphics*[width=65mm]{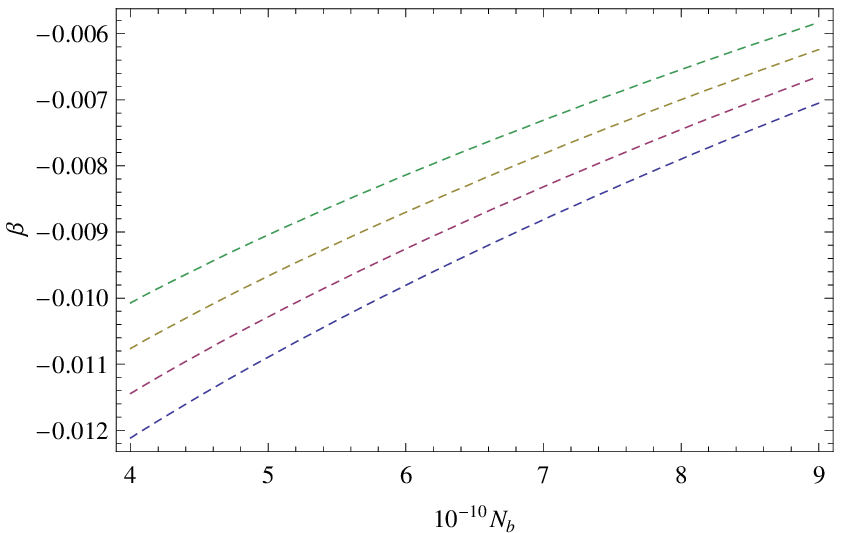}
  \caption{Analytical prediction of coefficient $\beta$ (\ref{quadratic_coeff}) for values $\delta_{max}\,=\,1.4\div2$ and $p\,=\,2$.}
  \label{plot_b_dmax}
\end{figure}

\begin{figure}[htbp]
  \centering
  \includegraphics*[width=65mm]{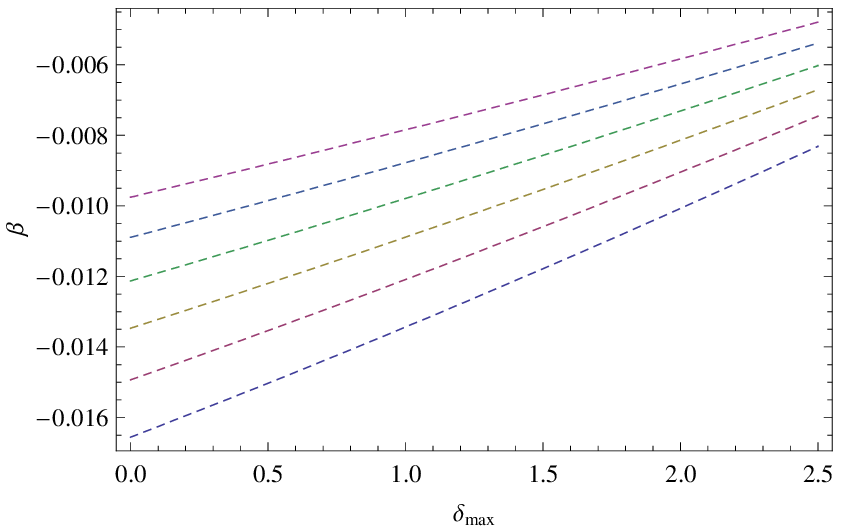}
  \caption{Analytical prediction of coefficient $\beta$ (\ref{quadratic_coeff}) for values $N_b\,=\,4\div9$ and $p\,=\,2$.}
  \label{plot_b_Nb}
\end{figure}

The coefficient $\beta$ can be found by
imposing the saturation condition of map (\ref{map}):
\begin{equation}\label{quadratic_coeff}
n_{sat}\,=\,\alpha\,n_{sat}+\beta\,{n_{sat}}^2\,\,\,\,\rightarrow\,\,\,\,\beta\,=\,\frac{1-\alpha}
{n_{sat}}
\end{equation}
and the map (\ref{map}) becomes
\begin{equation}\label{map_quadratic}
n_{m+1}\,=\,\alpha\,n_m+\frac{1-\alpha}{n_{sat}}\,{n_m}^2
\end{equation}
In Fig. (\ref{plot_b_dmax}), (\ref{plot_b_Nb}) we show the trends of the coefficient (\ref{quadratic_coeff}) as a function of $\delta_{max}$ for various values of bunch population and viceversa.

\section{Results and Conclusions}
\begin{figure}[htb]
  \centering
  \includegraphics*[width=65mm]{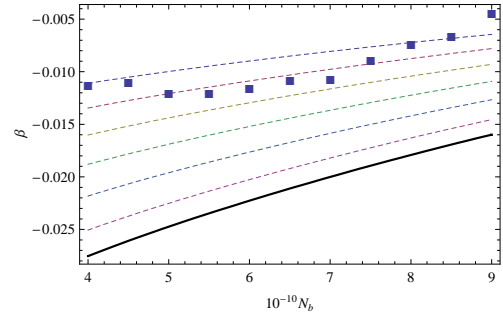}
  \caption{Comparison of the quadratic coefficient $\beta$ (Eq. (\ref{quadratic_coeff})) derived using ECLOUD simulations (points) and using the analysis of previous sections (dashed lines) with $p\,=\,2\,\div\,3$. The solid line is the result by assuming an uniform density.}
  \label{plot_b}
\end{figure}

In Figs. \ref{plot_b}   the analytical behavior and the outcomes of simulations (ECLOUD code) of  $\beta$  coefficient using the parameters reported  in Table \ref{tab1} show an acceptable agreement.
As a future work the analytical result could be useful to determine safe regions in parameter space where to minimize the  electron clouds.
Furthermore we would extend our results to include the presence of a  magnetic field. 

\end{document}